\documentclass[twocolumn,aip,showpacs,superscriptaddress,reprint]{revtex4-1}
\usepackage{mathrsfs,bm,multirow,amsmath,amsfonts,amssymb,array,booktabs,float}
\usepackage{graphicx,graphics,times,graphics,color,epsfig,epstopdf}
\renewcommand{\thispagestyle}[1]{}

\begin{document}

\title{A comprehensive first-principle study of borophene-based nano gas sensor with gold electrodes}

\author{Yueyue Tian}
\affiliation{International Centre for Quantum and Molecular Structures and Department of Physics, Shanghai University, Shanghai 200444, China}
\author{Houping Yang}
\affiliation{International Centre for Quantum and Molecular Structures and Department of Physics, Shanghai University, Shanghai 200444, China}
\author{Junjun Li}
\affiliation{Hongzhiwei Technology (Shanghai) Co., Ltd., Shanghai 201206, China}
\author{Shunbo Hu}
\affiliation{International Centre for Quantum and Molecular Structures and Department of Physics, Shanghai University, Shanghai 200444, China}
\affiliation{Materials Genome Institute, Shanghai University, Shanghai 200444, China}
\author{Shixun Cao}
\affiliation{International Centre for Quantum and Molecular Structures and Department of Physics, Shanghai University, Shanghai 200444, China}
\affiliation{Materials Genome Institute, Shanghai University, Shanghai 200444, China}
\author{Wei Ren}
\affiliation{International Centre for Quantum and Molecular Structures and Department of Physics, Shanghai University, Shanghai 200444, China}
\affiliation{Materials Genome Institute, Shanghai University, Shanghai 200444, China}
\author{Yin Wang}
\email{yinwang@shu.edu.cn}
\affiliation{International Centre for Quantum and Molecular Structures and Department of Physics, Shanghai University, Shanghai 200444, China}
\affiliation{Hongzhiwei Technology (Shanghai) Co., Ltd., Shanghai 201206, China}

\begin{abstract}
Using density functional theory combined with nonequilibrium Green's function method, the transport properties of borophene-based nano gas sensors with gold electrodes are calculated, and comprehensive understandings regarding the effects of gas molecules, MoS$_2$ substrate and gold electrodes to the transport properties of borophene are made. Results show that borophene-based sensors can be used to detect and distinguish CO, NO, NO$_2$ and NH$_3$ gas molecules, MoS$_2$ substrate leads to a non-linear behavior on the current-voltage characteristic, and gold electrodes provide charges to borophene and form a potential barrier, which reduced the current values compared to the current of the systems without gold electrodes. Our studies not only provide useful information on the computationally design of borophene-based gas sensors, but also help understand the transport behaviors and underlying physics of 2D metallic materials with metal electrodes.
\end{abstract}

\maketitle

Gas sensors have received considerable attentions because of their important applications in detecting toxic and deleterious gases, supervising air quality, and monitoring human health.\cite{LiuReview, YunusaReview} Compared to other gas detection techniques, electrochemical sensing has various advantages, such as, low energy linear output with high resolution, good selectivity and repeatability, high detection accuracy, and being more inexpensive than other techniques.\cite{YunusaReview} As motivated by the driving force of enhancing the performance of gas sensors, great efforts have been exhausted on their design with tiny architectures or configurations.\cite{LiuReview} At present, numerous distinct gas sensors have been prepared, for instance, Zhang \emph{et. al.}\cite{Zhang2014} have built a gas sensor based on anatase TiO$_{2}$ that is used to detect decomposed gases in SF$_{6}$, Cui \emph{et. al.}\cite{Cui2015} have constructed a field effect transistor sensor device which exhibits an ultrahigh sensitivity to NO$_{2}$ in dry air, Cho \emph{et. al.}\cite{Cho2015} have demonstrated atomically thin heterostructured gas sensors which comprising a combination of exfoliated MoS$_{2}$ and CVD graphene, Zhao \emph{et. al.}\cite{Zhao2018} have made a MoS$_{2}$-based flexible and wearable gas sensor which was fabricated based on the direct synthesis of MoS$_{2}$ on a polyimide substrate.

A clear trend is, with the emergence of two-dimensional (2D) materials,\cite{MaoR,WangR,HuangR} gas sensors based on 2D materials have drawn more and more attentions because of their superior properties, including large surface to volume ratio, associated charge transfer between gas and surface, and high electron mobility.\cite{Shukla2017, 2017Sc2CO2, APLMoSe2, APLZnO} Recently, graphene and other 2D materials-based gas sensors have been widely investigated both experimentally and theoretically.\cite{APLGraphene, lu2009gas, GGR, MalikR} Among these 2D materials, borophene, which has various excellent properties, such as anisotropic mechanical,\cite{wang2016strain, peng2017stability, zhang2017elasticity} optical\cite{gao2017prediction,peng2016electronic} and electronic properties,\cite{feng2016experimental, feng2016direct,wu2012two} plays a crucial role in promoting the development of 2D gas sensors.\cite{kou2014phosphorene,kootenaei2016b36, WangR} Shukla \emph{et. al.}\cite{Shukla2017} reported the realization of 2D monolayer borophene-based gas sensor using first principle density functional theory methods, and concluded that 2D monolayer borophene can be a potentially important substrate for gas sensing of CO, NO, NO$_2$ and NH$_3$ gas molecules. Furthermore, Shen \emph{et. al.}\cite{Shen2020} studied the transport properties of gas adsorbed on 2D monolayer borophene with a MoS$_2$ substrate, i.e., on borophene-MoS$_2$ heterostructure, and results show that the heterostructure is also very sensitive to CO, NO, NO$_2$ and NH$_3$ gas molecules, especially to NO moleclue, and therefore can be used as a 2D gas sensor. However, these studies ignored the effect of metal electrodes, which is necessary for not only the collecting of electrical signal during the operation of the sensor, but also adjusting the Fermi energy and/or electrostatic potential of the 2D materials.\cite{Chen1, Chen2, Xie} In addition, a comprehensive understanding on the effects of MoS$_2$ substrate, gas molecules, as well as metal electrodes, to the transport properties of 2D monolayer borophene is still lack.

In this paper, we systematically investigated the transport properties of borophene-based nano gas sensors with or without considering the influence of gas molecules, MoS$_2$ substrate and/or metal electrodes. Results show that the borophene and borophene-MoS$_2$ heterostructure with gold electrodes is sensitive to various gas molecules, therefore, they can be used as a gas sensor, which qualitatively agrees with previous studies.\cite{Shukla2017, Shen2020} The calculated current-voltage (\emph{I-V}) characteristic through pristine borophene without considering the effect of MoS$_2$ substrate and gold electrodes under 0.5 V has a linear behavior, however, with considering the MoS$_2$ substrate and/or the metal electrodes, the current is reduced and a non-linear \emph{I-V} behavior can be observed. Further band structure, charge and potential analyses well explained the obtained results. Our studies not only provided comprehensive information regarding the underlying physics of borophene-based gas sensors, but also can be used to understand more transport behaviors of 2D monolayer materials and heterostructures with metal electrodes, which are usually not included when theoretically studying the transport properties of 2D metallic materials in the literature.

\begin{figure}[tbp]
\includegraphics[width=7.5cm]{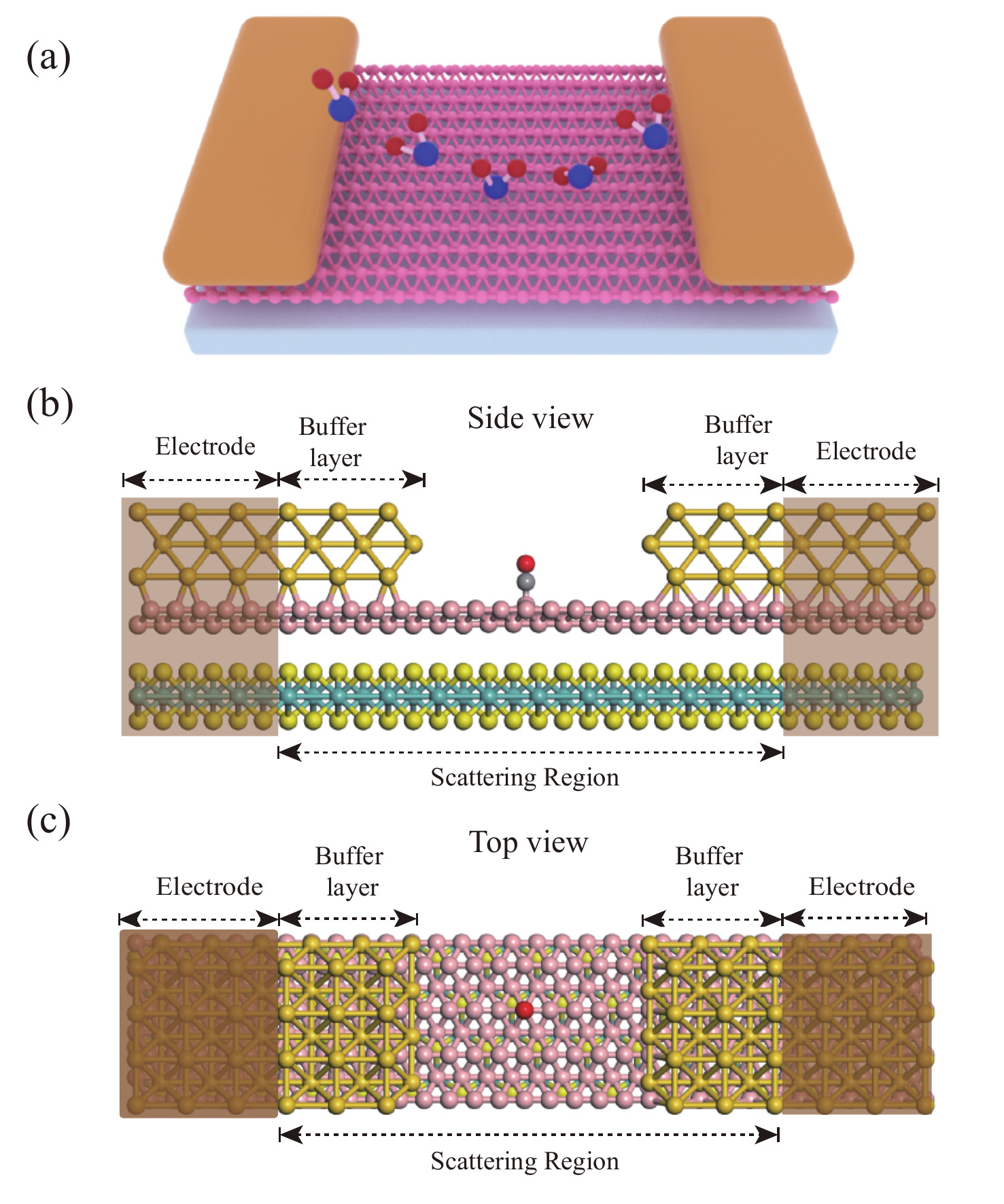}
\caption{(a) Schematic plot of the borophene-MoS$_2$ heterostructure gas sensor with gold electrodes. Gray area shows MoS$_2$ substrate, and gold areas are metal electrodes that collect the electrical signal. (b) Side view and (c) top view of the atomic model in our first-principle calculations. }\label{fig1}
\end{figure}

We started by considering the fabrication and operation process of a real borophene-based nano gas sensor with gold electrodes, as shown in Fig.~\ref{fig1}(a). Monolayer borophene is placed on MoS$_2$ substrate, and gold electrodes are deposited onto borophene. The distance between two gold electrodes is big enough to contain a gas molecule, so that the molecule can be regarded as no interaction with the electrodes. When a gas molecule is approaching and adsorbing on the borophene, the current is collected through the electrodes. Fig.~\ref{fig1}(b, c) show the side view and top view of the optimized atomic structure of the entire gas sensor. The distance between each two parts (borophene, MoS$_2$ or two gold electrodes) of the structure is optimized by \textsc{vasp} software,\cite{vasp} the optimization details can be found in the supplementary material. To calculate the transport properties without MoS$_2$ substrate and/or gold electrodes, we just need to simply remove the corresponding part from the entire structure in Fig.~\ref{fig1}. According to the previous studies,\cite{Shukla2017, Shen2020, Aghaei2018} in this work we considered four gas molecules, namely, CO, NO, NO$_2$ and NH$_3$, which are all sensitive to borophene-based gas sensors due to chemisorption between the molecules and borophene. These molecule-borophene structures are fully relaxed (see the supplementary material), while pristine borophene is used for all the systems without molecules. After the structure was determined, we carried the quantum transport calculations by density functional theory combined with the nonequilibrium Green's function (NEGF-DFT) as implemented in \textsc{nanodcal} software package.\cite{NEGFDFT1, NEGFDFT2} In our NEGF-DFT numerical calculations, double-zeta polarized atomic orbital basis set was used to expand the physical quantities of B and molecule elements, single-zeta polarized atomic orbital basis set was used for Mo, S and Au.\cite{basis} The exchange and correlation were treated at the level of the Perdew-Burke-Ernzerhof (PBE) level,\cite{PBE} and atomic cores are defined by the standard norm conserving nonlocal pseudopotentials.\cite{PP} In the self-consistent calculations, 3 k-points are used to sample the Brillouin zone in the periodic direction, and a cutoff energy of 80 Hartree were used. To obtain the transmission and \emph{I-V} curve, 50 k-points are used. Spin orbital coupling is not considered in this work.

\begin{figure}[tbp]
\includegraphics[width=7.5cm]{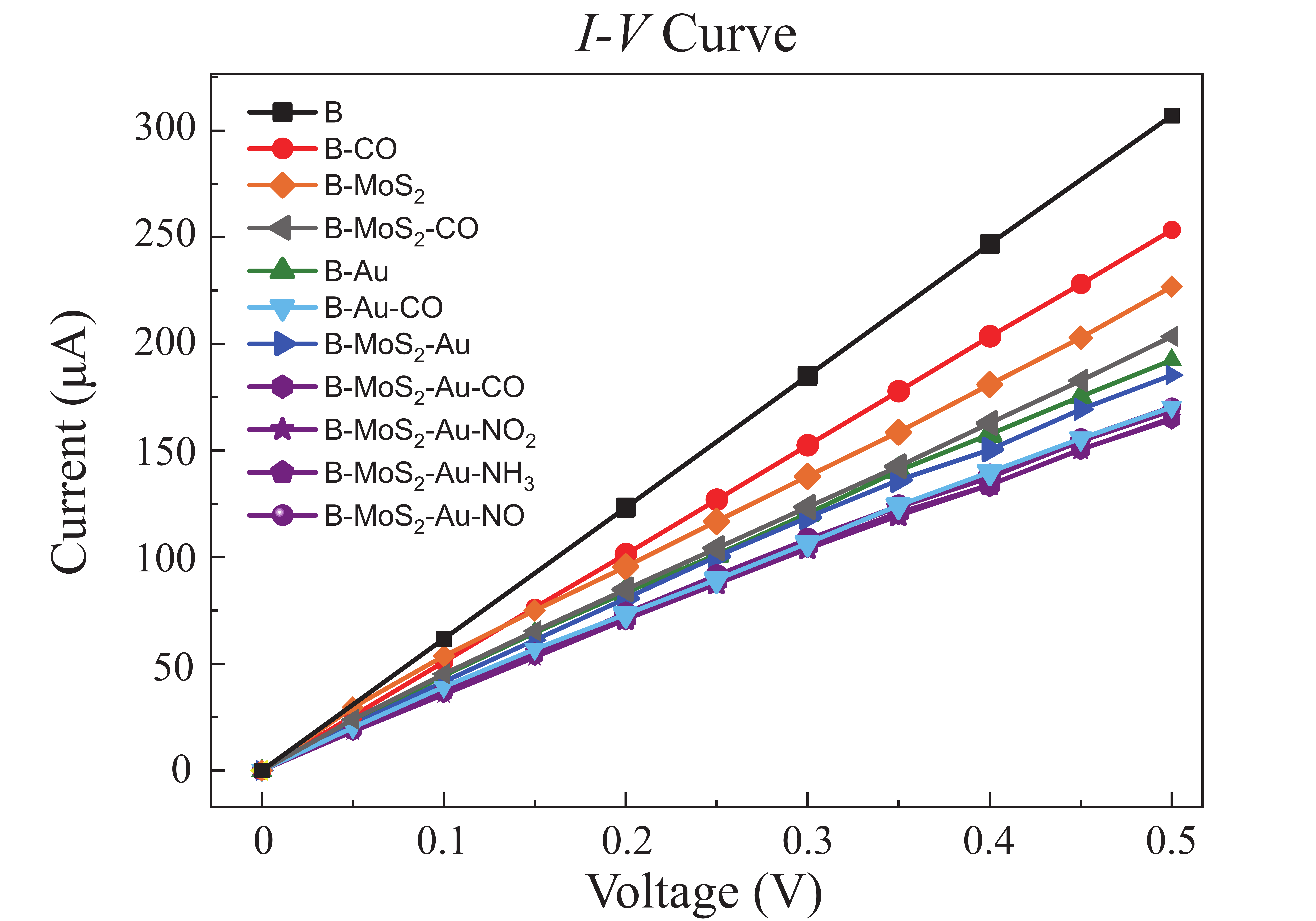}
\caption{Current-voltage (\emph{I-V}) characteristics of borophene-based nano gas sensors under 0.5 V external bias on various systems with or without considering the influence of gas molecules, MoS$_2$ substrate and/or gold electrodes.}\label{fig2}
\end{figure}

Fig.~\ref{fig2} shows the \emph{I-V} curves of borophene-based nano gas sensors under 0.5 V from our NEGF-DFT numerical calculations on eleven different systems with or without considering the influence of gas molecules, MoS$_2$ substrate and/or gold electrodes, from which the following results can be clearly observed:

(i) The \emph{I-V} characteristics of pristine borophene (black curve) show a nearly perfect linear behavior, which maintains after a molecule (CO) is adsorbed (red curve), and the scattering of the CO molecule leads to the decreasing of the current. Our calculated \emph{I-V} values quantitatively agree with the previous reports.\cite{Shukla2017} We should note that for the pristine borophene system (black) and borophene-MoS$_2$ heterostructure system (orange) that will be discussed in the following, the atomic structures are actually perfect periodic along the transport direction. For a periodic system, it is experimentally unfeasible to direct apply finite bias on a selected region, but numerical calculations of these periodic systems are very useful for one to better understand the physics of transport properties through the structure.

(ii) After MoS$_2$ is introduced, the current values of heterostructure systems (orange and gray curves) decrease compared to the current values of the corresponding borophene systems. More interestingly, under 0.15 V the \emph{I-V} characteristic show non-linear behavior, and then become nearly linear after 0.2 V, which is more obvious for the orange curve of borophene-MoS$_2$ system in Fig.~\ref{fig2}. This non-linear behavior was not mentioned in previous studies, and should be further understood. Quantitatively, our calculated current values under the same voltage is smaller than that in the previous reports on borophene-MoS$_2$ heterostructure,\cite{Shen2020} which is because we applied different strain on MoS$_2$ or borophene layer and used different stacking way to make the bilayer structure.

(iii) The rest seven \emph{I-V} curves are all for the systems with gold electrodes. We can see from the supplementary material that all these curves lose the perfect linear behavior, which must stem from the existence of metal electrodes. In addition, after gold electrodes are included, the calculated current values under the same voltage is much smaller than the current values of the corresponding system without gold electrodes, although gold is a good conductor from a macro point of view. Similar experimental observations are reported by Ref.~\onlinecite{ChoR} in the Al:Graphene system, where Al electrodes reduced the current of pure graphene.

(iv) It is obvious that for all the sensors, the current values of systems with molecules are smaller than that without molecules, which agrees with all the previous reports.\cite{Shukla2017, Shen2020} Moreover, one can see that the four purple curves quantitatively have nearly the same \emph{I-V} characteristic, which qualitatively agrees with the previous results on monolayer borophene gas sensor for CO, NO, NO$_2$ and NH$_3$ gas molecules.\cite{Shukla2017} Therefore, to further judge if the borophene-based nano-gas sensors can well distinguish these molecules, more studies on the sensitivities of the sensors to various molecules are necessary.

From the \emph{I-V} curves in Fig.~\ref{fig2}, we can preliminarily conclude that borophene-based nano heterostructure gas sensor with gold electrodes are sensitive enough to detect some small molecules. Till now, two behaviors observed in the \emph{I-V} curves (Fig.~\ref{fig2}) still need to be further understood. First, why MoS$_2$ substrate lead to a non-linear \emph{I-V} characteristic under small bias? Second, why gold electrode, which is a good conductor, lead to the decrease of current values compared to that in the corresponding systems without gold electrodes? In the following we will solve these two problems (See Fig. 5 and related discussions).

\begin{figure}[tbp]
\includegraphics[width=7.5cm]{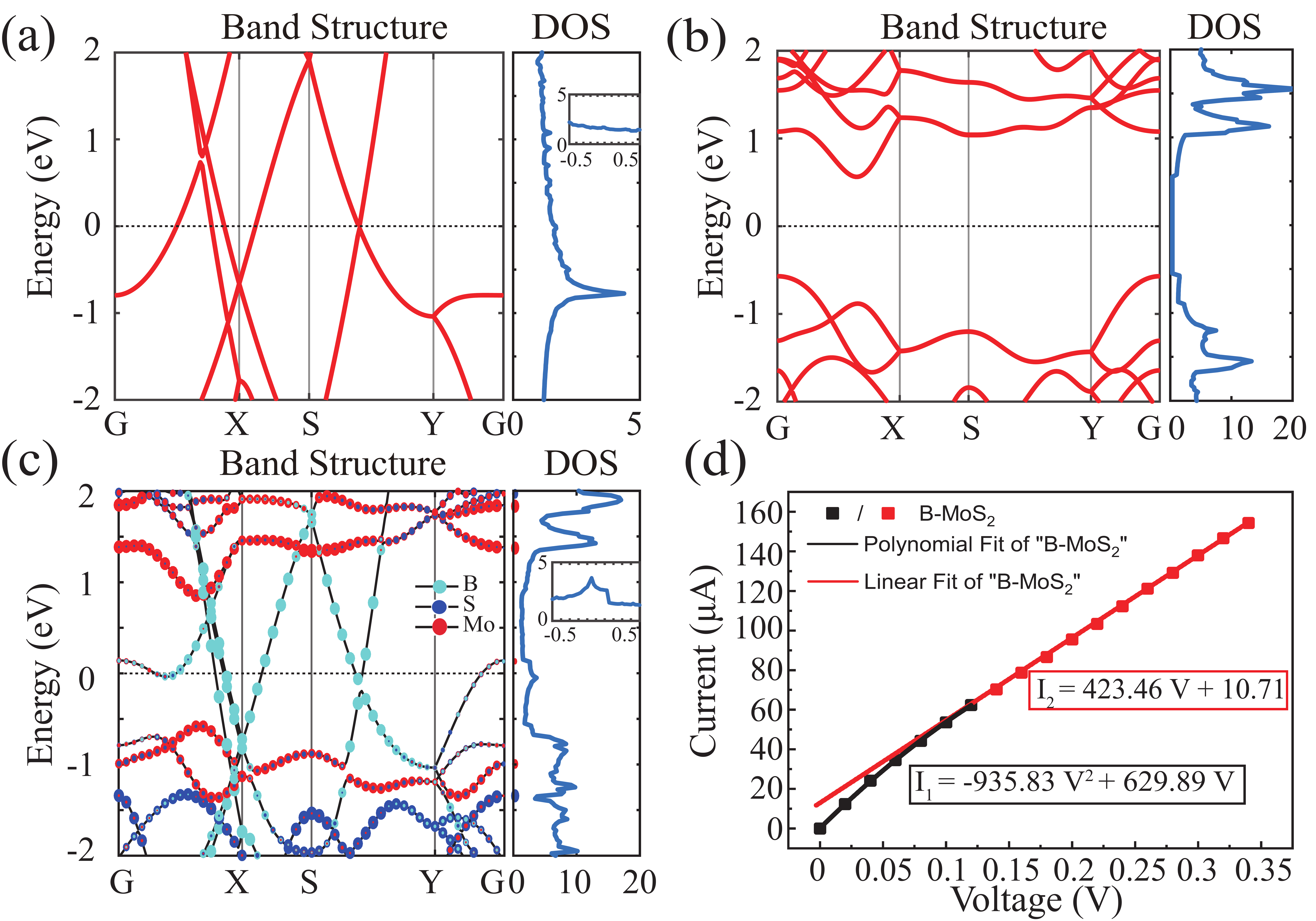}
\caption{Band structure and density of states (DOS) for (a) pristine borophene of a 8-atom unit cell system, (b) MoS$_2$ that matches the pristine borophene lattice, and (c) borophene-MoS$_2$ heterostructure. (d) The \emph{I-V} characteristic of borophene-MoS$_2$ system.}\label{fig3}
\end{figure}

Fig.~\ref{fig3}(a) shows the band structure and density of states (DOS) of pristine borophene calculated from 8-atom unit cell that matches MoS$_2$ lattice (see supplementary materials), one can observe that each energy band in the band structure cross the entire energy range from -0.5 to 0.5 eV in a nearly linear fashion, and the corresponding DOS nearly keep a constant value ($\sim$ 1.5). This means, if an external bias under 0.5 V is applied onto the borophene, each incoming electron state in the left electrode at a certain moment $k$ and energy $E$ can find a way out through and outgoing electron state in the right electrode at the same moment k with a different energy $E'$. In this way, the current will be proportional to the width of the bias widow, namely, the applied external bias voltage, leading to a linear behavior on the \emph{I-V} characteristic. After MoS$_2$, whose band structure and DOS is shown in Fig.~\ref{fig3} (b), and borophene forming a heterostructure, charge transfer will take place between these two materials. Fig.~\ref{fig3}(c) shows the band structure and DOS of borophene-MoS$_2$ heterostructure, and the contribution of each element is decomposed in the band structure. Clearly, in G-X and G-Y region above the Fermi energy, Mo element contributes to the energy band, due to the charge transfer from Mo to B (See later for charge discussion). Moreover, this energy band is in a non-linear fashion and ends at 0.143 eV, leading to an abrupt increase of the DOS ($\sim$ 4) above the Fermi energy. In this way, this energy band in both electrodes will contribute to the transport in a resonance way when the external bias is below 0.143 V. After the applied bias is over 0.143 V, this energy band in two electrodes will lose resonance, namely, the incoming electron state corresponding to the DOS peak in the left electrode miss the outgoing electron state corresponding to the DOS peak in the right electrode. We believe that the resonance of the DOS peak in both electrodes is the main reason that the \emph{I-V} characteristic of borophene-MoS$_2$ system has a non-linear behavior under smaller external bias. Fig.~\ref{fig3}(d) shows the \emph{I-V} curve of borophene-MoS$_2$ system with a smaller bias interval of 0.02 V, and a piecewise function is used to fit the curve as: $I_1=-935.83~V^2+629.89~V$ ($V\leq0.14$ V), and $I_2=423.46~V+10.71$ ($V\geq0.16$ V). The crossover point of these two fitted curves is at V=0.139 V, which is close to 0.143 V.

\begin{figure}[tbp]
\includegraphics[width=7.5cm]{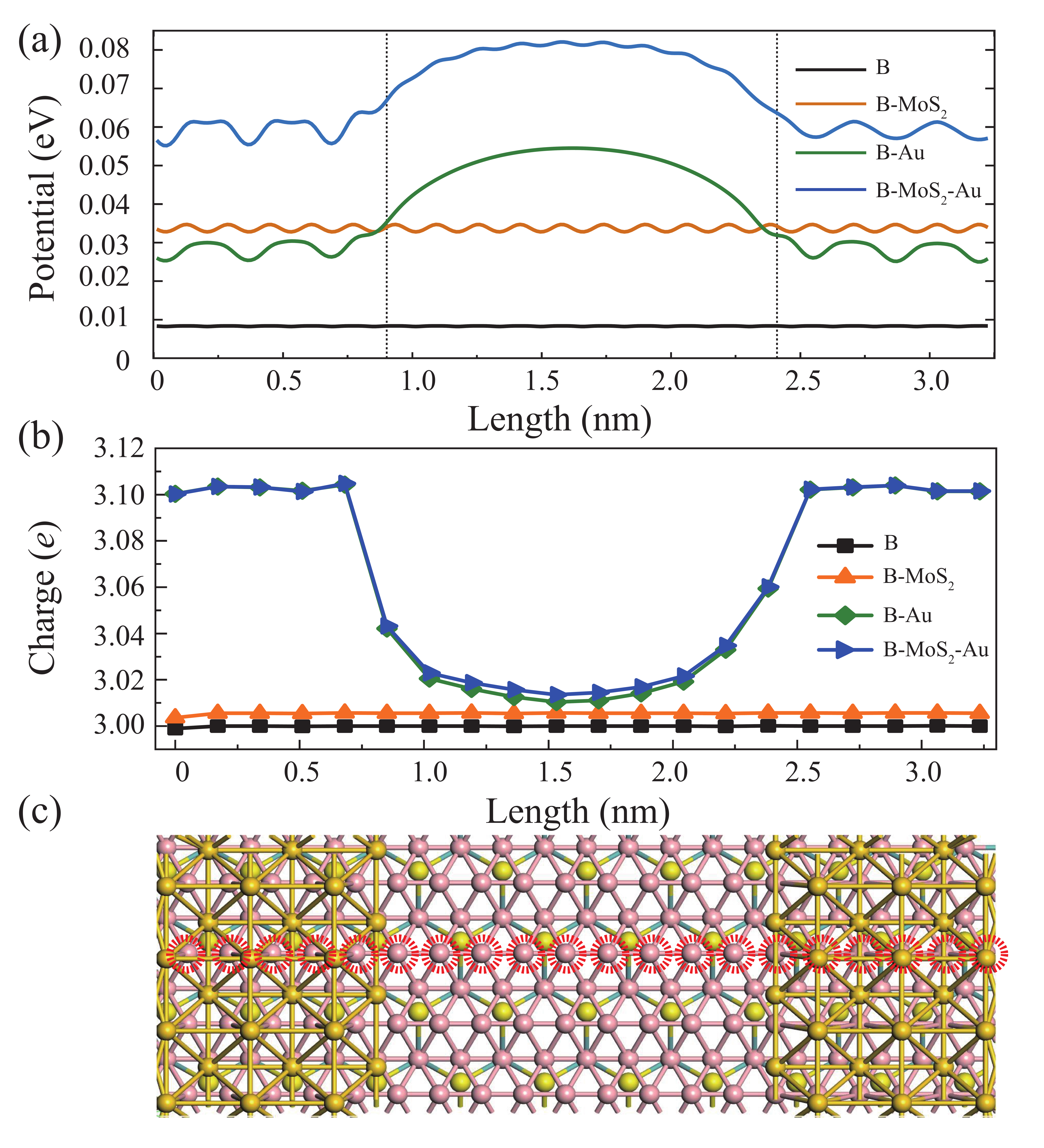}
\caption{(a) Electrostatic potential and (b) charge of borophene, borophene-MoS$_2$ with and without gold electrodes. (c) The charge of B atoms indicated by red dashed circles are calculated.}\label{fig4}
\end{figure}

We now turn to understand why currents of the systems with gold electrodes are drastically reduced compared to the current values of the systems without gold electrodes. Ref.~\onlinecite{Oliver} studied the ballistic conduction across the intermetallic interface between gold and tungsten, and concluded that the conduction is reduced compared to the conduction of pure gold or tungsten because of dissimilar electronic structure that gold has $s$ wave-like modes but tungsten has $d$ wave-like modes. Ref.~\cite{Wu} and ~\cite{Kang} also think that one factor to influence the spin injection efficiency from magnetic metal to graphene or transition-metal dichalcogenide semiconductors is orbital overlaps. In our case, borophene is actually extended along the transport direction and gold acts as top contacts\cite{Kang}, that is, the $p$ orbitals of borophene can contribute to the electron transport through the entire system, therefore, we still seek another understanding on the current reducing with gold electrodes.

Fig.~\ref{fig4}(a) shows the averaged electrostatic potential of four systems without molecules over the transverse direction. Clearly, for the pristine borophene and heterostructure systems, the potential is periodic along the transport direction, which is determined by the periodicity of the 2D crystal. For the systems with gold electrodes, a potential barrier is formed in the scattering region due to charge transfer. In this way, the incoming electron should pass through this potential barrier before it goes out through the right electrode. We also made a charge analysis as shown in Fig.~\ref{fig4}(b), in which the valence charge values of several B atoms in a selected row as indicted in Fig.~\ref{fig4}(c) is plotted. For pristine borophene the charge is +3 on all the B atoms. After MoS$_2$ substrate is added, electron transfer takes place from Mo atoms to B atoms, leading to slightly increase of the charge on B atoms, these charges contribute to the energy bands of the heterostructure around the Fermi energy as shown in Fig.~\ref{fig3}(c). After gold electrodes are introduced, more electrons transfer from the electrodes to B atoms, leading to potential barrier forming in the scattering region. Although gold is a good conductor and borophene also shows metallic state, this potential barrier hinders the electron transport and results in the decreasing of the current values.

\begin{figure}[tbp]
\includegraphics[width=7.5cm]{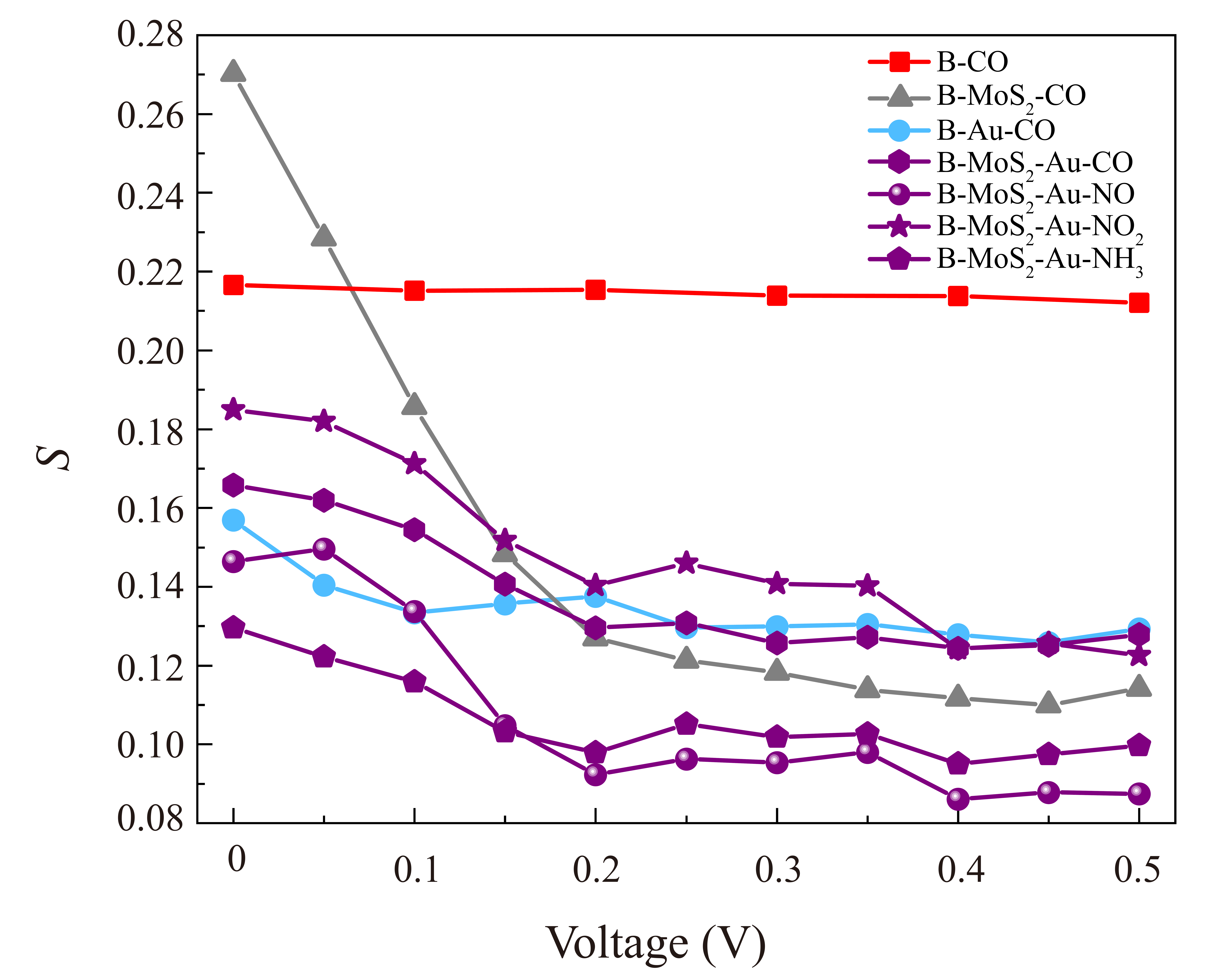}
\caption{The sensitivities of various borophene-based nano gas sensor to molecules.}\label{fig5}
\end{figure}

We finally calculated the sensitivity $S$ of the borophene-based nano gas sensor to various molecules (Fig.~\ref{fig5}), which is defined as $S=|I_0-I|/I$, where $I$ and $I_0$ are the currents with and without molecules, respectively.\cite{sensi, Shen2020} The sensitivity of the pristine borophene gas sensor keeps a high and nearly constant value (~0.22), the sensitivity of the borophene-MoS$_2$ heterostructure gas sensor quickly and smoothly reduced with increasing of the bias voltage. After gold electrodes are added, the sensitivities keep below 0.2. We should point that under 0.1 V bias voltage the sensitivities of heterostructure gas sensor with gold electrodes (purple curves) can be clearly distinguished, which means that the four molecules we studied in this paper can be well detected and differed by the gas sensor.

In summary, we studied the quantum transport properties of borophene-based nano gas sensors with or without considering the influence of gas molecules, MoS$_2$ substrate and/or metal electrodes using density functional theory combined with nonequilibrium Green's function method. Numerical results show that the gas sensors with gold electrodes can well detected and distinguished CO, NO, NO$_2$ and NH$_3$ gas molecules. The effects of MoS$_2$ substrate and gold electrodes to the transport properties of borophene are calculated and analyzed by electronic structure, charge, and electrostatic potential. Our studies provide useful information to understand the transport behaviors between 2D metallic materials with metal electrodes, which is usually not considered when theoretically investigating the transport properties of 2D metallic materials from first-principle in the literature.

See the supplementary material for (1) the optimized atomic details of borophene and MoS$_2$ heterostructure; (2) Relaxed structures of borophene monolayer after adsorption of four different gas molecules; (3) The \emph{I-V} curve of ten different systems; (4) transmission for eleven different systems with or without considering the influence of gas molecules, MoS2 substrate or gold electrodes at equilibrium state and 0.5 V bias voltage and the transmission versus moment k at Fermi energy.

\begin{acknowledgments}
Y.T. is grateful to Mr Guodong Zhao for useful discussions regarding the use of \textsc{nanodcal} software package. This work was financially supported by the National Key R$\&$D Program of China (Grant No. 2018YFB040760) and the National Natural Science Foundation of China (Grant No. 11774217). Y.T. and H.Y. were partially supported by the Postgraduate Research Opportunities Program of Hongzhiwei technology (Shanghai) Co., Ltd. (hzwtech-PROP).
\end{acknowledgments}


\end{document}